\documentclass[aps,pre,twocolumn,groupedaddress,showpacs,amsmath,amssymb]{revtex4}
\usepackage{epsfig}
\usepackage{amssymb}
\usepackage{amsmath}
\usepackage{graphicx}
\usepackage{amsfonts}
\usepackage{epsfig}
\usepackage{revsymb}
\usepackage{bm}
\usepackage{amsmath}
\usepackage{amssymb}
\usepackage{latexsym}
\usepackage{thumbpdf}
\usepackage{hyperref}

\begin{document}

\title{Charge-induced instability and macroscopic quantum nucleation phenomena
at the crystal $^4$He facet}

\author{S. N. Burmistrov}

\affiliation{Kurchatov Institute, 123182 Moscow, Russia}


\begin{abstract}
\par
An existence of the charge-induced instability is well known for the $^4$He
crystal surface in the rough state. Much less is known about charge-induced
instability at the $^4$He crystal surface in the smooth well-faceted state below
the roughening transition temperature. To meet the lack, we examine here the
latter case. As long as the electric field normal to the crystal facet is below
the critical value same as for the rough surface, the crystal faceting remains
absolutely stable. Above the critical field, unlike absolutely unstable state of
the rough surface, the crystal facet crosses over to the metastable state
separated from new crushed state with a potential barrier proportional to the
square of the linear facet step energy. The onset and development of the
instability at the charged crystal facet has much in common with the nucleation
kinetics of first-order phase transitions. Depending on the temperature, the
electric breaking strength is determined either by thermal activation at high
temperatures or by quantum tunneling at sufficiently low temperatures.

\end{abstract}


\maketitle

\section{Introduction}
\par
It is well known that a charged interface between two fluids can develop an
electrohydrodynamic instability at sufficiently high density of charges. Such
charge-induced instability results from the competition between the electric
forces of like charges and forces of surface tension and gravity. Naturally, the
liquid phases of helium have become one of physical systems for the theoretical
and experimental studies of electrohydrodynamic instabilities \cite{Lei92}, in
particular, as softening of the gravitational-capillary wave spectrum \cite{Gor,
Mim,Lei79}, charge-induced deformations of the interface \cite{Ebn,Ike,Shi},
formation of regular array of dimples \cite{Ebn,Ike,Shi}, individual multielectron
dimples \cite{Mel81}, and hexagonal reconstructuring \cite{Mel82}.
\par
It is of particular interest to compare the onset and development of the
electrohydrodynamic instability at the liquid-solid $^4$He interface with that at
the interface between two fluids. The first theoretical and experimental studies
have shown that a charged-induced instability at the superfluid-solid $^4$He looks
roughly like the electrohydrodynamic instability at the free liquid $^4$He surface
\cite{Sav,Uwa,Lei95, Bod}. For the high temperature region where the crystal
surface is in the rough state, such similar behavior is expectable since the
superfluid-solid interface in the rough state has a very high mobility and
interface excitations represent weakly damping crystallization waves whose
dispersion \cite{And,Kes} is quite similar to that of usual
gravitational-capillary waves at the free liquid surface.
\par
To date, no systematic study has been made on the onset and development of
electrohydrodynamic instability at the well-faceted and atomically smooth crystal
surfaces which may have an infinitely large stiffness and excitation spectrum
differing from the usual crystallization wave spectrum. The most striking
distinction of the smooth faceted crystal surface from the rough one is the
existence of nonanalytic cusplike behavior in the angle dependence for the surface
tension, e.g., \cite {Bal05,Lan,Noz}. The origin of the singularity is directly
connected with nonzero magnitude of the facet step energy below the roughening
transition temperature of about 1.2~K.
\par
In present work we attempt the electrohydrodynamic instability at the smooth
faceted surface of a $^4$He crystal in contact with its liquid phase. As we will
see below, the close similarity between the rough and smooth states of the crystal
surfaces extends until the charge density is below the critical one and the state
and shape of the surface are stable. As the charge density increases, the
development of the instabilities becomes different in kind. Unlike the rough
crystal surface, the faceted surface crosses over to a metastable state and the
further development of the instability is accompanied by overcoming some
nucleation barrier. The barrier height is proportional to the square of the facet
step energy and drastically reduces as the charge density increases. At the
sufficiently low temperatures the thermal activation mechanism of overcoming the
barrier is replaced with the quantum tunneling through the nucleation barrier. On
the whole, the charge-induced reconstructuring of the faceted crystal surface
resembles much first-order phase transitions and macroscopic quantum nucleation
phenomena in the helium systems \cite{Tsy,Ruu,Bur,Tan,Bal02}.
\par
For simplicity, we keep in mind the basal plane of hexagonal $^4$He crystal as an
example of the crystal facet and neglect any anisotropy in the plane. We also
suppose that the temperature is below the roughening transition temperature and
the crystal surface  is well-defined and faceted.

\section{Hamiltonian. The onset of instability at the crystal facet }
\par
Let us assume that the crystal surface is parallel to the $xy$ plane, with
vertical position at $z=0$. In order to consider the stability of the surface, we
proceed as follows. First, we call $\zeta =\zeta (\bm{r})$ the displacement of the
surface from its horizontal position $z=0$ with $\bm{r}=(x,y)$ as a
two-dimensional radius-vector. In addition to the surface tension force and the
force of gravity due to difference in the densities between the solid and liquid
states $\varDelta\rho$, one should involve also the interaction of the charges
with electric field $E$ and the direct Coulomb interaction between the charges.
Then the total energy $U$ of a charged surface can be written as
\begin{gather}
U=\int\! d^2r\biggl(\alpha (\bm{\nu})\sqrt{1+(\nabla\zeta )^2}+ \varDelta\rho\,
g\frac{\zeta ^2}{2} +eEn(\bm{r})\zeta\biggr)\nonumber
\\
+\frac{1}{2}\iint\! d^2r\, d^2r'\,\frac{en(\bm{r})en(\bm{r}')}{|\bm{l}-\bm{l}'|} .
\label{f01}
\end{gather}
Here $\bm{l}=(\bm{r},\zeta)$ stands for the three-dimensional coordinate of a
point at the surface, $n(\bm{r})$ is the density of electrons with charge $e$, $g$
is the acceleration of gravity, and $\alpha (\bm{\nu})$ is the energy of a unit
surface area or surface tension.
\par
Unlike the fluid-fluid interface, the surface tension coefficient $\alpha
(\bm{\nu})$ for the crystal facet depends essentially on the direction of the
normal $\bm{\nu}$ to the interface against crystallographic axes. In our simplest
case this is a function of angle $\vartheta$ alone between the normal and the
crystallographic [0001] or $c$-axis of the crystal hcp structure with the
geometric relation $\mid\tan\vartheta\mid=\mid\nabla\zeta\mid$.
\par
For the crystal facet tilted by small angle $\vartheta$ from the basal plane, the
expansion of surface tension $\alpha (\vartheta )$ usually written, e.g.
\cite{Bal05,Lan,Noz}, as
\begin{equation*}
\alpha (\vartheta ) = (\alpha _0+\alpha _1\tan\mid\vartheta\mid +
\ldots)\cos\vartheta , \;\;\;\;\;\; \mid\tan\vartheta\mid =\mid\nabla\zeta\mid ,
\end{equation*}
can be represented for the small angles by a series
\begin{equation*}
\alpha (\vartheta ) = \alpha (0)+\alpha _1\mid\vartheta\mid+ \alpha
^{\prime\prime}(0)\frac{\vartheta ^2}{2} +\ldots ,    \;\;\;\;\;\;
\mid\vartheta\mid\ll 1.
\end{equation*}
We intentionally do not write the next terms of expansion, e.g., cubic one due to
step-step interaction, since we assume to study only small bending of the crystal
surface. The angular behavior has a nonanalytic cusplike behavior at $\vartheta
=0$ due to $\alpha _1=\alpha _1(T)$ representing a ratio of the linear facet step
energy $\beta$ to the crystallographic interplane spacing. Below the roughening
transition temperature for the basal plane $T_R\sim$1.2~K the facet step energy
$\beta=\beta (T)$ is positive and vanishes for temperatures  $T>T_R$.
\par
To determine the equilibrium shape of the surface $\zeta (\bm{r})$ and equilibrium
charge distribution $n(\bm{r})$, we must minimize the energy functional
(\ref{f01}) against $\zeta$ and $n$ at a given total surface charge $Q$. This
condition can readily be taken into account by augmenting the energy functional
with the Lagrange factor $\lambda$ in the form
 $$
-\lambda \int en(\bm{r})\,d^2r .
 $$
In addition, treating the energy functional, we naturally imply one more obvious
condition $n(\bm{r})\geqslant 0$.
\par
In the general form the minimization of the energy functional is a practically
unsolvable problem because its expression (\ref{f01}) contains not only quadratic
terms in $\zeta$ and $n$. Thus, we start first from analyzing small surface
bending $\zeta (\bm{r})$ and small gradients $|\nabla\zeta |\ll 1$. The latter
implies implicitly that $|\zeta |\ll r$ and we can put approximately
$\bm{l}=\bm{r}$ in the denominator of the Coulomb term in Eq.~(\ref{f01}). Next,
we expand the surface tension term in $|\nabla\zeta|$, retain the linear and
quadratic terms alone, and arrive at the following expression for the total excess
energy $U$ associated with nonzero surface bending $\zeta (\bm{r})$
\begin{gather*}
U=\int\! d^2r\biggl(\alpha _1|\nabla\zeta|+\frac{\alpha _0}{2}(\nabla\zeta )^2+
\frac{\varDelta\rho\, g}{2}\zeta ^2 +eEn(\bm{r})\zeta \biggr)\nonumber
\\
+\frac{1}{2}\iint\! d^2r\, d^2r'\,\frac{en(\bm{r})en(\bm{r}')}{|\bm{r}-\bm{r}'|} .
\end{gather*}
Here we have labeled
 $$ \alpha _0=\alpha (0)+\alpha ^{\prime\prime}(0) $$
as a surface stiffness.
\par
The spatial scale of surface distortion is usually determined by the capillary
length $\lambda _0 =(\alpha _0/\varDelta\rho\, g)^{1/2}\approx 1$~mm if one takes
$\alpha _0\approx 0.2$~erg/cm$^2$ and $\varDelta\rho\approx 0.018$~g/cm$^3$ for
$^4$He \cite{Bal05}. The electric field $E$ and charge surface density $en$ have
the same dimensionality and their typical scale is $(\alpha _0\varDelta\rho\,
g)^{1/4}\approx 400$~V/cm. Accordingly, the typical electron density equals
$(\alpha _0\varDelta\rho\, g)^{1/4}/e\approx 2.8\times 10^{9}$~cm$^{-2}$. The
number of electrons $\pi\alpha _0^{5/4}(\varDelta\rho\, g)^{-3/4}$ within the
circle of radius $\lambda _0$ runs to $10^8$. And lastly, unit of energy
corresponds to $\alpha _0^2/(\varDelta\rho\, g)\approx 2.2\times 10^{-3}$~erg.
\par
So, if we measure $\zeta$ and $\bm{r}$ in units of capillary length $\lambda _0$,
electric field and charge density in units of $(\alpha _0\varDelta\rho\,
g)^{1/4}$, and energy in units of $\alpha _0^2/(\varDelta\rho\, g)$, the total
excess energy $U$ can be expressed in terms of dimensionless units as
\begin{gather}
U=\int\! d^2r\biggl(\frac{\alpha _1}{\alpha _0}|\nabla\zeta|+\frac{(\nabla\zeta
)^2}{2}+ \frac{\zeta ^2}{2} + En (\bm{r} )\zeta\biggr)\nonumber
\\
+\frac{1}{2}\iint\! d^2r\, d^2r'\,\frac{n(\bm{r})n(\bm{r}')}{|\bm{r}-\bm{r}'|} .
\label{f06}
\end{gather}
As for the step energy $\alpha _1$, we assume that its low temperature value
\cite{Bal05} is approximately $\alpha _1\approx 0.014$~erg/cm$^2$. This value
amounts to one-tenth of surface stiffness $\alpha _0$ and in the following we can
keep inequality $\alpha _1/\alpha _0\ll 1$ in mind. Moreover, this small parameter
justifies all approximations that will be made further.
\par
The uniform state of the surface holds for the electric field values as long as
the contribution to the excess energy (\ref{f06}) due to variations $\zeta
(\bm{r})$ from $\zeta =0$ and $n(\bm{r})$ from homogeneous value $\bar{n}$  is a
positive-definite quantity. Using the following equation for Lagrange factor
 $$
E\zeta (\bm{r})+\int\frac{n(\bm{r}')\, d^2r'}{|\bm{r}-\bm{r}'|}=\lambda
 $$
and putting $\delta n(\bm{r})=n(\bm{r})-\bar{n}$, we find for the variation of the
excess energy
\begin{gather}
\delta U=\int\! d^2r\biggl(\frac{\alpha _1}{\alpha
_0}|\nabla\zeta|+\frac{(\nabla\zeta )^2}{2}+ \frac{\zeta ^2}{2} + E\,\delta n
(\bm{r} )\zeta\biggr)\nonumber
\\
+\frac{1}{2}\iint\! d^2r\, d^2r'\,\frac{\delta n(\bm{r})\delta
n(\bm{r}')}{|\bm{r}-\bm{r}'|} . \label{f07}
\end{gather}
To analyze it, we use the Fourier representation
 $$
\zeta (\bm{r})=\sum\limits _{\bm{k}}\zeta
_{\bm{k}}e^{i{\bm{kr}}}\;\;\text{and}\;\; \delta n (\bm{r})=\sum\limits
_{\bm{k}}\delta n _{\bm{k}}e^{i{\bm{kr}}} ,
 $$
and rewrite the energy variation as
\begin{gather*}
\delta U=\int\! d^2r\, \frac{\alpha _1}{\alpha _0}|\nabla\zeta|+
\\
\frac{1}{2}\sum\limits _{\bm{k}}\bigl[ (k^2+1)\zeta _{\bm{k}}\zeta _{\bm{k}}^{*}+
E(\delta n_{\bm{k}}\zeta _{\bm{k}}^{*}+\delta n_{\bm{k}}^{*}\zeta
_{\bm{k}})+\frac{2\pi}{k}\delta n_{\bm{k}}\delta n_{\bm{k}}^{*}\bigr] .
\end{gather*}
Minimizing $\delta U$ over $\delta n _{\bm{k}}$ yields the optimum relation
\begin{equation}\label{f10}
\delta n _{\bm{k}}=-\frac{k}{2\pi}\zeta _{\bm{k}}
\end{equation}
and the corresponding optimum value of the energy
\begin{equation}\label{f11}
\delta U=\int\frac{\alpha _1}{\alpha _0}|\nabla\zeta|\, d^2r+
\frac{1}{2}\sum\limits _{\bm{k}}\bigl(k^2+1-\frac{kE^2}{2\pi}\bigr)|\zeta
_{\bm{k}}|^2 .
\end{equation}
The second term is always positive provided the inequality $E^2<2\pi(k+1/k)$ is
satisfied for all wave vectors $k$. The minimum of the right-hand side of the
inequality occurs at $k=k_c=1$ and corresponds to the critical field
$E_c=\sqrt{4\pi}$. Thus, the crystal facet is absolutely stable at $E<E_c$.
\par
At $E>E_c$ the stability is lost and the distortions of the homogeneous state
should appear. In this regard the situation resembles the loss of stability for
the rough state of the crystal surface. However, the development of the stability
and the transition to unhomogeneous state differ drastically. In fact, due to
positive $\alpha _1>0$ term linear in $|\nabla\zeta |$ the evolution of the
crystal facet perturbations should inevitably be accompanied with overcoming some
potential barrier, the barrier height being dependent on the field strength $E$.
The more the field strength, the less the potential barrier height.
\par
To proceed, let us return to the coordinate representation of Eq.~(\ref{f11})
\begin{gather}
\delta U=\int\! d^2r\biggl(\frac{\alpha _1}{\alpha
_0}|\nabla\zeta|+\frac{(\nabla\zeta )^2}{2}+ \frac{\zeta ^2}{2}\biggr)\nonumber
\\
-\frac{1}{2}\iint\! d^2r\, d^2r'\,\frac{E^2}{(2\pi )^2}\frac{\bigl(\nabla
_{\bm{r}}\zeta (\bm{r})\nabla _{\bm{r}'}\zeta (\bm{r}')\bigr)}{|\bm{r}-\bm{r}'|}
\label{f12}
\end{gather}
and give a qualitative description of the matter. For this purpose, we employ a
variational principle and dimensional analysis of the functional (\ref{f12}). Let
us represent the surface distortion $\zeta (\bm{r})$ with the aid of the trial
function $f(x)$ in the axially symmetrical form as
\begin{equation}\label{f13}
\zeta (\bm{r})=\zeta f(r/R),
\end{equation}
where $\zeta$ is a typical magnitude of distortion and $R$ is its typical size.
Then we have
\begin{equation*}\label{f14}
\delta U(\zeta , R)=\frac{\alpha _1}{\alpha _0}a|\zeta |R +b\frac{\zeta
^2}{2}+c\frac{\zeta ^2R^2}{2}-d\frac{E^2}{4\pi ^2}\frac{\zeta ^2R}{2} ,
\end{equation*}
and the dimensionless factors are given by
\begin{gather*}
a=\int _0^{\infty}|f'(r)|2\pi r\, dr, \;\;\; b=\int _0^{\infty}f^{\prime\,
2}(r)2\pi r\, dr,
\\
c=\int _0^{\infty}f^2(r)2\pi r\, dr,\;\;\; d=\int d^2r\, d^2r'\frac{\bm{r\cdot
r}'}{rr'}\frac{f'(r)f'(r')}{|\bm{r}-\bm{r}'|}
\\
= \int _0^{\infty}dk\biggl(\int _0^{\infty}dr\, 2\pi rf'(r)J_1(kr)\biggr)^2 ,
\end{gather*}
where $J_1(x)$ is the Bessel function of the first kind.
\par
As the electric field strength exceeds the value $E_0=(8\pi ^2\sqrt{bc}/d)^{1/2}$,
there appears a region of $\zeta$ and $R$ with the negative values of $\delta U$
separated always from $\delta U=0$ at $\zeta =0$ with the intermediate positive
$\delta U$ values.  Rewriting the excess energy $\delta U(\zeta , R)$ as
\begin{gather*}\label{f15}
\delta U=\frac{1}{2}\biggl(\frac{8\pi^2\alpha _1}{\alpha
_0}\biggr)^2\frac{a^2b/d^2}{E^4-E_0^4} +\frac{b}{2}\zeta ^2
\biggl(1-\sqrt{\frac{c}{b}}\,\frac{E^2}{E_0^2}R\biggr)^2
\\
-\frac{c}{2}\biggl(\frac{E^2}{E_0^2}-1\biggr) \biggl(|\zeta |R-\frac{\alpha
_1}{\alpha _0}\frac{a/c}{E^4/E_0^4 -1}\biggr)^2, \; E_0^2=\frac{8\pi
^2\sqrt{bc}}{d},
\end{gather*}
one can readily see that the state of the crystal facet changes from the stable to
metastable state at $E>E_0$ and the point
\begin{equation*}\label{f16}
|\zeta _0|=\frac{\alpha
_1}{\alpha_0}\frac{a}{\sqrt{bc}}\,\frac{E^2E_0^2}{E^4-E_0^4}
\;\;\;\text{and}\;\;\; R_0=\sqrt{\frac{b}{c}}\,\frac{E_0^2}{E^2}
\end{equation*}
becomes a saddle point of the potential relief. The potential barrier height equal
to
\begin{equation*}\label{f17}
U_0= \frac{a^2}{2c}\,\frac{\alpha _1^2}{\alpha _0^2}\, \frac{E_0^4}{E^4-E_0^4}
\end{equation*}
must be overcome to break the flat faceting of a crystal surface.
\par
Unfortunately, we cannot find the exact function $f(r)$ and, correspondingly,
values of factors $a$, $b$, $c$ and $d$ which optimize the functional (\ref{f12}).
However, it is clear that the potential barrier height should be infinitely large
at $E=E_c$ and thus $E_0=E_c$ for the exact solution. This entails the obvious
relation $d=2\pi (bc)^{1/2}$ between coefficients for the exact solution. To
estimate them, we use a trial function $f(x)=\exp (-x^2)$. The direct calculation
results in
\begin{equation*}\label{f18}
a=\pi ^{3/2},\;\; b=\pi ,\;\; c=\pi /2,\;\; d=\pi ^{5/2}/\sqrt{2},
\end{equation*}
and
\begin{equation}\label{f18a}
\frac{E_c}{E_0}=\sqrt{\frac{d}{2\pi\sqrt{bc}}}=\frac{\pi ^{1/4}}{\sqrt{2}}\approx
0.94
\end{equation} in place
of unity. Hence we may expect an accuracy of our estimate within about 10\%.
\par
Let us compare the height $U_0$ of the potential barrier at the saddle point with
the roughening transition temperature $T_R$ about 1.2~K. In the dimensional units
we have
\begin{equation*}\label{f19}
U_0=\frac{a}{2c}\frac{\alpha _1^2}{\varDelta\rho\, g}\frac{E_c^4}{E^4-E_c^4}\sim
1.4\times 10^{11}\frac{E_c^4}{E^4-E_c^4}\;\; (\text{in K}).
\end{equation*}
One may be surprised with the huge barrier height so that, unlike the rough
crystal surface absolutely unstable at $E\geqslant E_c$, tens of $E_c$ should keep
a crystal facet practically stable for an experimentally available time. Provided
we expect a reasonable observation time of destructing the faceted state due to
thermal activation mechanism, we should provide a ratio $U_0/T$ of about a few
tens \cite{Bur,Tan} This means that the electric field $E$ should exceed the
critical one $E_c$ by a factor of about 300. The same factor certainly refers to
the surface density of charges.
\par
In the dimensional CGSE units the bending deflection $\zeta _0$ and the typical
size of inhomogeneity $R_0$ are given by
\begin{gather*}
|\zeta _0| =4\pi\frac{a}{\sqrt{bc}}\,\alpha _1\frac{E^2}{E^4-E_c^4}\sim
31\frac{\alpha _1E^2}{E^4-E_c^4} ,
\\
R_0 =\sqrt{\frac{b}{c}}\,\frac{4\pi\alpha _0}{E^2}\sim 18\frac{\alpha _0}{E^2}.
\label{f20}
\end{gather*}
In the weak fields of few critical values the critical parameters $R_0$ and
$|\zeta _0|$ prove to be of macroscopic sizes in accordance with macroscopically
large height of the potential barrier.
\par
For $E=300E_c$, we find approximately $R_0\sim 16$~nm and $|\zeta _0|\sim 2$~nm.
On the whole, the electric field should be very large compared with the critical
value $E_c$ in order to reduce significantly the nucleation barrier for the
effective production of a few circular crystal terraces tilted with the angle
about $\arctan (\alpha _1/\alpha _0)\sim 4^{\circ}$. In this sense the critical
fluctuation represents a region of the crystal surface in the rough state.
\par
From the physical point of view the angle of slope $\arctan (\alpha _1/\alpha
_0)\sim 4^{\circ}$ is determined by the competition of two contributions into the
total surface energy. One originates from the regular surface term $\alpha _0\zeta
^2$ and the second does from irregular step tension term $\alpha _1|\zeta |R$.
Provided $\alpha _0\zeta ^2\gg\alpha _1|\zeta |R$, the latter contribution becomes
negligible and thus the interface properties resemble those in the rough surface
state. On the contrary, if $\alpha _0\zeta ^2\ll\alpha _1|\zeta |R$, the dominant
term linear in $|\zeta |$ is responsible for the origin of a potential barrier
since the other terms quadratic in $\zeta$ are yet insignificant.
\par
Note that the small gradient approximation we use is satisfied since $|\nabla\zeta
|\sim |\zeta _0|/R_0\sim \alpha _1/\alpha _0\ll 1$ with the exception of narrow
region $E\sim E_c$. The latter remark refers also to justifying small density
variations $\delta n \ll\bar{n}$ valid to the extent of smallness $|\zeta
_0|/R_0$.

\section{Lagrangian. The quantum breaking of the crystal facet}
\par
The destruction of the faceted crystal surface is accompanied by overcoming some
potential barrier depending on the charge surface density. There are two basic
mechanisms to overcome the potential barrier. One is the thermal activation
efficient at high temperatures and the second is the quantum tunneling through a
potential barrier dominant at sufficiently low temperatures. In order to treat the
quantum tunneling, it is necessary to involve the interface dynamics, in
particular, to determine the kinetic energy of the charged interface in addition
to the potential energy $U$.
\par
As a first step, we employ the so-called metallic approximation. In this
approximation it is assumed that the mobility of electrons along the
superfluid-crystal He$^4$ interface is very high and the charged helium interface
represents an equipotential surface so that the electric field is always normal to
the interface as for a well-conducting metal. A necessary condition for such
approximation assumes at least that the plasma oscillation frequency of a
two-dimensional layer of electrons with effective mass $m_e$
\begin{equation}\label{f20a}
\Omega _p\sim (2\pi ne^2k/m_e)^{1/2}
\end{equation}
is much larger than the typical frequency $\omega$ of the gravitational-capillary
or melting-crystallization waves at the same wave vector $k$. So, within our first
approximation we believe that the charge density distribution $n(t,\bm{r})$ has
sufficient time to accommodate to the surface distortion $\zeta (t,\bm{r})$ and is
determined by the electrostatic relations in accordance with the profile $\zeta
(t,\bm{r})$.
\par
Neglecting possible energy dissipation, we describe the charged interface dynamics
using the following action
\begin{equation}\label{f21}
S=\int dt\, L[\zeta (t,\,\bm{r}),\dot{\zeta} (t,\,\bm{r}),n(t,\bm{r})]
\end{equation}
with the Lagrangian $L$ equal to the difference between the kinetic energy
functional and the potential energy functional $U$ introduced by Eq.~(\ref{f01})
\begin{equation*}\label{f22}
 L=\frac{\rho _\text{eff}}{2}\!\iint\! d^2r\,
d^2r'\,\frac{\dot{\zeta}(t,\,\bm{r})\dot{\zeta}(t,\,\bm{r}')}{2\pi
|\bm{r}-\bm{r}'|} - U[\zeta(t,\,\bm{r}), n(t,\,\bm{r})] .
\end{equation*}
Here we ignore the compressibility of the both liquid and solid phases. Because of
low temperature consideration we will also neglect the normal component density in
the superfluid phase or, equivalently, difference between the superfluid density
$\rho _s$ and the density of the liquid phase $\rho$. Then the effective interface
density $\rho _\text{eff}$ is given by
\begin{equation*}
\rho _\text{eff}=(\rho '-\rho)^2/\rho \approx 1.9\,\text{mg/cm}^3
\end{equation*}
and depends on the difference  $\varDelta\rho =\rho '-\rho$ between the solid
density $\rho '$ and the liquid density $\rho$. For our purposes, the exact
magnitude of the effective density is inessential.
\par
Next, for convenience, let us introduce units of time equal to $(\rho
_{\text{eff}}\lambda _0^3/\alpha _0)^{1/2}\approx 3.1$~ms and measure the action
in units of $(\alpha _0\rho _{\text{eff}}\lambda _0^7)^{1/2}\approx 0.62\times
10^{-5}$~erg$\cdot$s. Using the speculations and arguments bringing us to
Eq.~(\ref{f10}) and then to Eq.~(\ref{f12}), we arrive at examining the following
effective action
\begin{equation*}\label{f23}
S=\int dt\, L_{\text{eff}}[\zeta (t,\,\bm{r}),\dot{\zeta} (t,\,\bm{r})]
\end{equation*}
with the dimensionless Lagrangian
\begin{gather*}\label{f23a}
 L_{\text{eff}}=\frac{1}{2}\!\iint\! d^2r\,
d^2r'\,\frac{\dot{\zeta}(t,\,\bm{r})\dot{\zeta}(t,\,\bm{r}')}{2\pi
|\bm{r}-\bm{r}'|}
\\
-\int\! d^2r\biggl(\frac{\alpha _1}{\alpha _0}|\nabla\zeta|+\frac{(\nabla\zeta
)^2}{2}+ \frac{\zeta ^2}{2}\biggr)\nonumber
\\
+\frac{1}{2}\iint\! d^2r\, d^2r'\,\frac{E^2}{(2\pi )^2}\frac{\bigl(\nabla
_{\bm{r}}\zeta (t,\bm{r})\nabla _{\bm{r}'}\zeta
(t,\bm{r}')\bigr)}{|\bm{r}-\bm{r}'|}.
\end{gather*}
\par
Within an exponential accuracy the quantum decay rate of the metastable state is
proportional to
\begin{equation*}\label{f24}
\Gamma\propto\exp (-S_E/\hbar ),
\end{equation*}
where $S_E$ is the effective Euclidean action calculated at the optimum escape
path. This path starts at the entrance point under the potential barrier and ends
at the point at which the optimum fluctuation escapes from the barrier
\cite{Bur,Tan}. In other words, quantum fluctuation penetrates through the
potential barrier along the path of least resistance. Before calculating the
quantum rate at which the crystal facet breaks up, we must go over to the
effective Euclidean action defined in imaginary time $t\rightarrow it$. We refer
to books \cite{Qua,Wei} for details.
\par
As a result, we should analyze the following functional defined within the time
interval $[-\hbar/2T,\,\hbar/2T]$
\begin{gather*}\label{f25}
S_E=\int dt\, L_E[\zeta (t,\,\bm{r}),\dot{\zeta} (t,\,\bm{r})],
\\
 L_E=\frac{1}{2}\!\iint\! d^2r\,
d^2r'\,\frac{\dot{\zeta}(t,\,\bm{r})\dot{\zeta}(t,\,\bm{r}')}{2\pi
|\bm{r}-\bm{r}'|}
\\
+\int\! d^2r\biggl(\frac{\alpha _1}{\alpha _0}|\nabla\zeta|+\frac{(\nabla\zeta
)^2}{2}+ \frac{\zeta ^2}{2}\biggr)\nonumber
\\
-\frac{1}{2}\iint\! d^2r\, d^2r'\,\frac{E^2}{(2\pi )^2}\frac{\bigl(\nabla
_{\bm{r}}\zeta (t,\bm{r})\nabla _{\bm{r}'}\zeta
(t,\bm{r}')\bigr)}{|\bm{r}-\bm{r}'|}.
\end{gather*}
Again, the exact determination of extrema for the action $S_E$ is a rather
complicated problem. We here consider only the case of zero temperature when the
limits of integration over imaginary time are infinite. As before, it is
convenient to take an advantage of the dimensional analysis and variational
principle. We will express the surface distortion $\zeta (t,\bm{r})$ in the terms
of function $f(y,x)$ with the scaled arguments as
\begin{equation*}\label{f26}
\zeta (t,\bm{r})=\zeta f(t/\tau , r/R) .
\end{equation*}
Next, we calculate the action $S_E$ at zero temperature
\begin{gather}
S_E(\zeta ,\tau , R)= F\frac{\zeta ^2R^3}{2\tau}+ \tau\biggl(\frac{\alpha
_1}{\alpha _0}A|\zeta |R +B\frac{\zeta ^2}{2}\nonumber
\\
+C\frac{\zeta ^2R^2}{2}-D\frac{E^2}{4\pi ^2}\frac{\zeta ^2R}{2}\biggr).
\label{f27}
\end{gather}
The numerical factors are given by the integrals
\begin{gather*}\label{f28}
A=\int _{-\infty}^{\infty}dt\int _0^{\infty}|f'(t,r)|2\pi r\, dr,
\\
B=\int _{-\infty}^{\infty}dt\int _0^{\infty}f^{\prime\, 2}(t,r)2\pi r\, dr,
\\
C=\int _{-\infty}^{\infty}dt\int _0^{\infty}f^2(t,r)2\pi r\, dr,
\\
D=\int  _{-\infty}^{\infty}dt\int d^2r\, d^2r'\frac{\bm{r\cdot
r}'}{rr'}\frac{f'(t,r)f'(t,r')}{|\bm{r}-\bm{r}'|}
\\
= \int _{-\infty}^{\infty}dt\int\limits _0^{\infty}dk\biggl(\int _0^{\infty}dr\,
2\pi rf'(t,r)J_1(kr)\biggr)^2 ,
\\
F=\int _{-\infty}^{\infty}dt\int d^2r\,
d^2r'\frac{\dot{f}(t,r)\dot{f}(t,r')}{2\pi|\bm{r}-\bm{r}'|}
\\
= \int _{-\infty}^{\infty}dt\int\limits _0^{\infty}\frac{dk}{2\pi}\biggl(\int
_0^{\infty}dr\, 2\pi r\dot{f}(t,r)J_0(kr)\biggr)^2 ,
\end{gather*}
where $J_0(x)$ and $J_1(x)$ are the Bessel function of the first kind.
\par
From the condition of vanishing derivatives in $\zeta$, $R$ and $\tau$ for $S_E$
we find the following parameters of the quantum critical fluctuation
\begin{gather*}\label{f29}
|\zeta _q|=\frac{A}{\sqrt{BC}}\frac{\alpha _1}{\alpha
_0}\biggl(\sqrt{1-\frac{7}{16}\frac{E_0^4}{E^4}}
+\frac{3}{4}\biggr)\frac{E^2E_0^2}{E^4-E_0^4},
\\
R_q=\frac{7}{4}\sqrt{\frac{B}{C}}\frac{E_0^2}{E^2}
\biggl(\sqrt{1-\frac{7}{16}\frac{E_0^4}{E^4}}+1\biggr)^{-1},
\\
\tau
_q=\frac{7}{2\sqrt{2}}\biggl(\frac{F^2B}{C^3}\biggr)^{1/4}\!\!\frac{E_0^3}{E\sqrt{E^4-E_0^4}}
\frac{\bigl(\sqrt{1-\frac{7}{16}\frac{E_0^4}{E^4}}+\frac{3}{4}\bigr)^{1/2}}
{\sqrt{1-\frac{7}{16}\frac{E_0^4}{E^4}} +1}.
\end{gather*}
Here $E_0=(8\pi ^2\sqrt{BC}/D)^{1/2}$ which should coincide for the exact solution
with the critical field value, i.e., $E_0=E_c=\sqrt{4\pi}$. Then we calculate the
corresponding value of action $S_q$ according to
\begin{equation*}\label{f29a}
S_q=\frac{A}{2}\frac{\alpha _1}{\alpha _0}|\zeta _q|R_q\tau _q
\end{equation*}
at the critical point $(\zeta _q, R_q, \tau _q)$ representing a saddle point of
the functional $S_E$ (\ref{f27}). Finally, we obtain
\begin{gather*}\label{f30}
S_q=\frac{49A^2}{16\sqrt{2}}\frac{(BF^2)^{1/4}}{C^{7/4}}\frac{\alpha _1^2}{\alpha
_0^2}\frac{\bigl(\sqrt{1-\frac{7}{16}\frac{E_0^4}{E^4}}+\frac{3}{4}\bigr)^{3/2}}
{\bigl(\sqrt{1-\frac{7}{16}\frac{E_0^4}{E^4}}+1\bigr)^2}
\\
\times\frac{E_0}{E}\biggl(\frac{E_0^4}{E^4-E_0^4}\biggr)^{3/2}.
\end{gather*}
Like the potential barrier height, the action $S_q$ becomes infinite at the same
critical field $E=E_0$.
\par
To estimate the numerical coefficients $F$, $A$, $B$, $C$, and $D$, we choose a
physically expedient trial function $f(t,r)=\exp [-(t^2+r^2)]$. The
straightforward calculation gives
\begin{gather*}\label{f31}
A=\pi ^2,\; B=\frac{\pi ^{3/2}}{2^{1/2}},\; C=\biggl(\frac{\pi}{2}\biggr)^{3/2},\;
D=\frac{\pi^3}{2},\; F=\frac{\pi ^2}{4}
\end{gather*}
with the same ratio $E_c/E_0$ as in (\ref{f18a}).
\par
Let us compare the action $S_q$ with the Planck constant $\hbar$. Introducing a
facet capillary length $\lambda _1=(\alpha _1/\varDelta\rho\, g)^{1/2}$, we have
in the dimensional units
\begin{gather*}\label{f32}
\frac{S_q}{\hbar}=\frac{49A^2}{16\sqrt{2}}\frac{(BF^2)^{1/4}}{C^{7/4}}
\biggl(\frac{\sqrt{\alpha _1\alpha _0}\rho_{\text{eff}}\lambda _1^7}{\hbar
^2}\biggr)^{1/2}
\\
\times\frac{\bigl(\sqrt{1-\frac{7}{16}\frac{E_c^4}{E^4}}+\frac{3}{4}\bigr)^{3/2}}
{\bigl(\sqrt{1-\frac{7}{16}\frac{E_c^4}{E^4}}+1\bigr)^2}
\frac{E_c}{E}\biggl(\frac{E_c^4}{E^4-E_c^4}\biggr)^{3/2}
\\
\approx 4\times
10^{21}\frac{\bigl(\sqrt{1-\frac{7}{16}\frac{E_c^4}{E^4}}+\frac{3}{4}\bigr)^{3/2}}
{\bigl(\sqrt{1-\frac{7}{16}\frac{E_c^4}{E^4}}+1\bigr)^2}
\frac{E_c}{E}\biggl(\frac{E_c^4}{E^4-E_c^4}\biggr)^{3/2} .
\end{gather*}
As is seen, even for the electric fields which are dozens of times larger than the
critical one $E_c$, the ratio $S_q/\hbar$ has a giant magnitude so that the
crystal surface will remain in the well-defined faceted state for the practically
infinite time. At $E\gg E_c$ we have an estimate
\begin{equation*}\label{f33}
S_q/\hbar\approx 2.3\times 10^{21}(E_c/E)^7.
\end{equation*}
For strong $E\gg E_c$ fields, in the dimensional CGSE units the bending deflection
$\zeta _0$ and the typical size of inhomogeneity $R_0$ are given by
\begin{gather*}\label{f34}
|\zeta _q| =4\pi\frac{7A}{4\sqrt{BC}}\frac{\alpha _1}{E^2}\sim 80\frac{\alpha
_1}{E^2} ,
\\
R_q =\frac{7}{8}\sqrt{\frac{B}{C}}\,\frac{4\pi\alpha _0}{E^2}\sim 15\frac{\alpha
_0}{E^2}.
\end{gather*}
For $E=300E_c$, we find approximately $R_q\sim 13$~nm and $|\zeta _q|\sim 5$~nm.
The estimate of the tunneling time in the strong $E\gg E_c$ fields yields
\begin{gather*}\label{f35}
\tau _q\approx 10^{-2}(E_c/E)^3\;\; (\text{in seconds}).
\end{gather*}
\par
Again, the small gradient approximation is fulfilled since $|\nabla\zeta |\sim
|\zeta _q|/R_q\sim\alpha _1/\alpha _0\ll 1$. Let us compare the plasmon frequency
$\Omega _p$ with the inverse time of tunneling $\tau _q^{-1}$ in order to justify
the metallic approximation. We consider the case of strong fields and take $k\sim
1/R_q$ as a typical wave vector for the spatial size of the surface distortion.
Then, using (\ref{f20a}),
\begin{equation*}\label{f35a}
\Omega _p\tau _q\sim\sqrt{\frac{\rho
_{\text{eff}}}{\varDelta\rho}}\sqrt{\frac{eE_c}{m_eg}}\biggl(\frac{E_c}{E}\biggr)^{3/2}
\sim 2\times 10^{7}\biggl(\frac{E_c}{E}\biggr)^{3/2}.
\end{equation*}
Thus, in the fields $E=300E_c$ the fulfillment of inequality $\Omega _p\gg\tau
_q^{-1}$ evidences for the favor of the metallic approximation.

\section{Thermal-quantum crossover temperature. The decay rate}
\par
Let us turn to the thermal-quantum crossover temperature $T_q$ which separates the
classical thermal activation at $T>T_q$ from the quantum nucleation mechanism at
lower $T<T_q$ temperatures. Here we estimate the thermal-quantum crossover
temperature $T_q=T_q(E)$ as a ratio of the potential barrier height to the saddle
value $S_q$ of the Euclidean action at zero temperature. In the dimensional units
we have then
\begin{gather*}\label{f36}
T_q(E)=\frac{\hbar U_0}{S_q}=\hbar\biggl(\frac{\alpha _0}{\rho
_{\text{eff}}\lambda
_0^3}\biggr)^{1/2}\frac{8\sqrt{2}}{49}\frac{a^2C^{7/4}}{cA^2(BF^2)^{1/4}}
\\
\times\frac{\bigl(\sqrt{1-\frac{7}{16}\frac{E_c^4}{E^4}}+1\bigr)^2}
{\bigl(\sqrt{1-\frac{7}{16}\frac{E_c^4}{E^4}}+\frac{3}{4}\bigr)^{3/2}}
\frac{E\sqrt{E^4-E_c^4}}{E_c^3}\approx
\\
1.7\times 10^{-7}\frac{\bigl(\sqrt{1-\frac{7}{16}\frac{E_c^4}{E^4}}+1\bigr)^2}
{\bigl(\sqrt{1-\frac{7}{16}\frac{E_c^4}{E^4}}+\frac{3}{4}\bigr)^{3/2}}
\frac{E\sqrt{E^4-E_c^4}}{E_c^3} \; (\text{in mK}).
\end{gather*}
Note that the thermal-quantum crossover temperature is independent of the step
tension coefficient $\alpha _1$. This point is obvious since the barrier height
$U_0$ and action $S_q$ are both proportional to the same factor $\alpha _1^2$.
\par
At the electric fields comparable with the critical one $E_c$ the thermal-quantum
crossover temperature, starting from its zero value at $E=E_c$,  proves to be
extremely small. In the strong $E\gg E_c$ fields the thermal-quantum crossover
temperature grows approximately as a cube of the field
\begin{equation*}\label{f37}
T_q(E)\approx 3\times 10^{-7}(E/E_c)^3\;\;\; (\text{in mK}).
\end{equation*}
For fields $E=300E_c$, we may expect a reasonable magnitude for the
thermal-quantum crossover temperature of about 8~mK.
\par
Let us consider a charged crystal facet prepared in the metastable $E>E_c$ state
with adjusting thermodynamic parameters such as temperature $T$ and electric field
$E$. After the lapse of some time $t_{obs}$, there will appear a nucleus of the
rough state breaking the crystal faceting. Then the nucleation rate $\Gamma
=\Gamma (T,E)$ and the time of observation $t_{obs}$ are connected by the
following relation
\begin{equation*}\label{f38}
t_{obs}N_{nuc}\Gamma \simeq 1 ,
\end{equation*}
where $N_{nuc}$ is the total number of independent nucleation sites and $\Gamma$
is the nucleation rate at a single nucleation site. We estimate $N_{nuc}$
approximately as the total number of atoms at the crystal surface, assuming that
every atom at the surface has an equal possibility to become a nucleation site
within the time interval $t_{obs}$. For the crystal area of 1~cm$^2$, we put
\begin{equation*}\label{39}
N_{nuc}\sim 10^{14} .
\end{equation*}
\par
The nucleation rate $\Gamma$ can approximately be estimated as
\begin{equation*}\label{f40}
\Gamma\sim\nu\exp(-S)
\end{equation*}
where $\nu$ is the attempt frequency and exponent $S$, depending on temperature,
is either Arrhenius exponent $U_0/T$ or Euclidean one $S_q/\hbar$. The attempt
frequency $\nu$ is associated with the surface fluctuations resulting in nonzero
bending $\zeta (t,\bm{r})$ of the flat crystal facet. In general, the frequency of
crystal surface fluctuations depends on the magnitude of surface bending $\zeta$
and the radius of deformation $R$ as well. This frequency  can be estimated by
equating the kinetic energy to the potential surface energy in Lagrangian $L$
(\ref{f21}). The order-of-magnitude estimate can be represented as \cite{Bur11}
\begin{equation*}\label{41}
\nu\sim\biggl(\frac{\alpha _1R+\alpha _0|\zeta |}{\rho _{\text{eff}}|\zeta
|R^3}\biggr)^{1/2}.
\end{equation*}
\par
According to \cite{Sch}, there is one optimum path, i.e., escape path which
connects the entrance point with the optimum escape point and corresponds to the
saddle-point value of the effective Euclidean action. In the quasiclassical
approximation the main contribution to the decay rate of the metastable state is
determined by such optimum escape path and its nearest vicinity. As is found
above, at the optimum escape path a ratio of surface deformation $\zeta$ to its
radius $R$ satisfies approximately $|\zeta |/R\sim\alpha _1/\alpha _0$. Then we
arrive at
\begin{equation*}\label{42}
\nu\sim\biggl(\frac{\alpha _1^3}{\alpha _0^2\rho _{\text{eff}}|\zeta
|^3}\biggr)^{1/2}.
\end{equation*}
\par
Next, we should estimate the equilibrium fluctuations of the surface bending as a
function of temperature. At high temperatures one expects the thermal activation
mechanism when the average energy fluctuations should be of the order of the
temperature, i.e., $\alpha _1R|\zeta|+\alpha _0\zeta ^2\sim\alpha _0\zeta ^2\sim
T$. Hence, for $T=1$~K, we expect
\begin{equation*}\label{f43}
|\zeta |\sim 0.3\,\text{nm},\;\;\; R\sim 3\,\text{nm}\;\;\text{and}\;\;\nu\sim
5\times 10^{10}\,\text{Hz}.
\end{equation*}
At zero temperature the attempt frequency can be associated with the zero-point
oscillations in the same potential $U=\alpha _1R|\zeta|+\alpha _0\zeta
^2\sim\alpha _0\zeta ^2$. Using $U\sim\hbar\nu (U)$ for an estimate of the ground
level energy, we find
\begin{equation*}\label{f44}
\nu\sim\biggl(\frac{\alpha _1^6}{\alpha _0\hbar ^3\rho _{\text{eff}}^2}
\biggr)^{1/7}\sim\!\! 7\times 10^{10}\,\text{Hz},\;\;\;\; U=\hbar\nu\sim
0.5\,\text{K}.
\end{equation*}
\par
Note that the magnitude of the surface bending is about of the interatomic spacing
and the frequency has numerically the same order of the magnitude as the Debye
frequency. These magnitudes seem us reasonable. Thus, we have a relatively large
preexponential factor
\begin{equation*}\label{f45}
\nu N_{nuc}\sim 10^{24}\,\text{s}^{-1}\sim e^{55}\,\text{s}^{-1}
\end{equation*}
which can readily be compensated by macroscopically large potential barrier for
insufficiently high density of charges. Eventually, if we wish to discover the
process of the facet destruction for the time of about tens seconds, the exponents
$U_0/T$ or $S_q/\hbar$ should be kept about 55.
\par
Due to strong exponential dependence of nucleation rate $\Gamma$ on the
thermodynamic parameters $T$ and $E$ the statistical dispersion of nucleation
events is not large as compared with the average values of the thermodynamic
parameters at which the nucleation is mainly observed. The overwhelming majority
of experimental points will concentrate in the narrow region around the average
values which correspond to the so-called rapid nucleation line. In essence, from
the viewpoint of the time of observation  the rapid nucleation line separates the
metastable states into two region. One region represents the long-living states
looking as stable during the experiment and the other is the short-living states
which decay practically instantly.
\par
So, for the rapid nucleation line or the breaking field $E_b$, we may expect the
following behavior. Under thermal activation mechanism at high temperatures one
should observe
\begin{equation*}\label{f46}
E_b(T)\propto T^{-1/4},\;\;\; T>T_q.
\end{equation*}
Below the thermal-quantum crossover temperature this behavior should go over to
the practically temperature-independent behavior
\begin{equation*}\label{f47}
E_b(T)\approx\text{const},\;\;\; T<T_q.
\end{equation*}
\par
In the latter connection we would like to mention a possible effect of the energy
dissipation processes. As is known from the quantum dynamics of first-order phase
transitions \cite{Bur87}, the energy dissipation processes increase the effective
Euclidean action and thus reduce the quantum decay rate. Accordingly, the behavior
of the breaking field $E_b(T)$ in the quantum regime should grow with the
temperature rise and demonstrate a maximum at the thermal-quantum crossover
temperature. However, as is mentioned above, the energy dissipation in superfluid
$^4$He is not large at low temperatures because of negligible density of the
normal component. That is why, we expect only a slight manifestation of the energy
dissipation effects in the quantum regime.

\section{Summary}
\par
To summarize, we have examined a stability of the charged crystal $^4$He surface
in the atomically smooth and well-faceted state below the roughening transition
temperature. Like the charged crystal $^4$He surface in the rough state, the
charged crystal $^4$He facet becomes unstable at the same density of charges or
corresponding critical electric field $E_c$. However, the dynamics of the
transition from the initial homogeneous distribution of charges and flat crystal
surface to a spatially unhomogeneous charge distribution and to a warped crystal
surface proves to be qualitatively different.
\par
In the rough surface state, as the electric field exceeds the critical value
$E_c$, the homogeneous surface state becomes absolutely unstable and in this sense
the development of the charge-induced instability resembles a second-order phase
transition. In contrast, as the electric field exceeds the same critical value
$E_c$, the homogeneous state of the atomically smooth and well-faceted crystal
surface is converted into the metastable state separated with a potential barrier
governed by the electric field or charge density. The barrier height is
proportional to the square of the linear facet step energy.
\par
The onset and development of the charge-induced instability at the crystal facet
can be compared with the kinetics of first-order phase transitions accompanied by
the nucleation and next growth of new stable phase. A nucleus of new phase here
can be described as a fluctuation region of the crystal surface in the atomically
rough state. The larger the charge density, the smaller the radius of the critical
nucleus.
\par
Unlike the charged rough crystal surface, in the electric fields which are tens
times larger than the critical value $E_c$ the potential barrier still remains so
high that the charged crystal facet, though metastable, will not break up for the
experimentally obtainable uptime. To realize a breakage of the crystal facet, a
few hundreds of critical value $E_c$ should be achieved. The breaking electric
strength $E_b$ depends on the temperature as well as the nature of the breaking
mechanism. At high temperatures the breaking dynamics is associated with the
thermal activation mechanism and with the quantum tunneling through a potential
barrier at sufficiently low temperatures.

\section*{ACKNOWLEDMENTS}
\par
The author is thankful to V.~L.~Tsymbalenko for stimulating discussion. The work
is supported in part by the RFBR Grant No.~10-02-00047a.

\end{document}